\documentclass[conference,final]{IEEEtran}
\IEEEoverridecommandlockouts

\usepackage[utf8]{inputenc}
\usepackage[T1]{fontenc}
\usepackage{microtype,newtxtext,newtxmath} 
\usepackage{mathtools,gensymb,eurosym,url,soul} 
\usepackage{subcaption}
\usepackage{longtable,booktabs,array}
\usepackage[backend=biber, sorting=none, style=ieee, maxnames=5]{biblatex}
\usepackage[switch]{lineno}
\usepackage{acronym}
\usepackage{graphicx}
\usepackage{xcolor}
\usepackage{xr} 
\usepackage{textcomp}
\usepackage{hyperref}
\hypersetup{colorlinks=true, breaklinks=true, 
  urlcolor=blue, linkcolor=blue,anchorcolor=blue,citecolor=blue,
  hypertexnames=true, final=true, 
  pdfpagemode = UseNone, 
  pdfauthor = {},
  pdftitle = {},   
  pdfsubject = {},
  pdfkeywords = {}
}

\graphicspath{{figures/} {img/} {./}}
\DeclareGraphicsExtensions{.pdf,.png,.jpg,.mps}

\DeclareSourcemap{
  \maps[datatype=bibtex]{
    \map[overwrite=true]{
      \step[fieldset=url, null]
      \step[fieldset=issn, null]
      \step[fieldset=doi, null]
      \step[fieldset=eprint, null]
    }
  }
}

\IEEEdisplaynontitleabstractindextext
\begin{document}

\title{Core interface optimization for multi-core neuromorphic processors\\
{\footnotesize \textsuperscript{}}
\thanks{}
}

\author{\IEEEauthorblockN{Zhe Su}
\IEEEauthorblockA{\textit{Institute of Neuroinformatics} \\
\textit{University of Zurich and ETH Zurich}\\
Zurich, Switzerland \\
zhesu@ini.uzh.ch}
\and
\IEEEauthorblockN{Hyunjung Hwang}
\IEEEauthorblockA{\textit{ETH Zurich} \\
Zurich, Switzerland \\
hhwang@student.ethz.ch}
\and
\IEEEauthorblockN{Tristan Torchet}
\IEEEauthorblockA{\textit{ETH Zurich} \\
Zurich, Switzerland \\
ttorchet@ethz.ch}
\and
\IEEEauthorblockN{Giacomo Indiveri}
\IEEEauthorblockA{\textit{Institute of Neuroinformatics} \\
\textit{University of Zurich and ETH Zurich}\\
Zurich, Switzerland \\
giacomo@ini.uzh.ch}

\thanks{This work was partially supported by the European Research Council (ERC) under the European Union’s Horizon 2020 Research and Innovation Program Grant Agreement No. 724295 (NeuroAgents), and by the Electronic Component Systems for European Leadership (ECSEL) joint undertaking Grant Agreement No. 876925 (ANDANTE).}
}

\maketitle

\begin{abstract}
Hardware implementations of Spiking Neural Networks (SNNs) represent a promising approach to edge-computing for applications that require low-power and low-latency, and which cannot resort to external cloud-based computing services. However, most solutions proposed so far either support only relatively small networks, or take up significant hardware resources, to implement large networks. To realize large-scale and scalable SNNs it is necessary to develop an efficient asynchronous communication and routing fabric that enables the design of multi-core architectures. In particular the core interface that manages inter-core spike communication is a crucial component as it represents the bottleneck of Power-Performance-Area (PPA) especially for the arbitration architecture and the routing memory. In this paper we present an arbitration mechanism with the corresponding asynchronous encoding pipeline circuits, based on hierarchical arbiter trees. The proposed scheme reduces the latency by more than 70\% in sparse-event mode, compared to the state-of-the-art arbitration architectures, with lower area cost. The routing memory makes use of asynchronous Content Addressable Memory (CAM) with Current Sensing Completion Detection (CSCD), which saves approximately 46\% energy, and achieves a 40\% increase in throughput against conventional asynchronous CAM using configurable delay lines, at the cost of only a slight increase in area. In addition as it radically reduces the core interface resources in multi-core neuromorphic processors, the arbitration architecture and CAM architecture we propose can be also applied to a wide range of general asynchronous circuits and systems.
\end{abstract}


\section{Introduction}
Neuromorphic processors are event-based processing architectures that adopt in-memory computing strategies and brain-inspired principles of computation to implement computational models of Spiking Neural Networks (SNNs)~\cite{Chicca_etal14}.
Due to their asynchronous and spike-based data-driven processing nature, they have the potential of achieving ultra-low power computations for edge-computing applications. 
An efficient way to build large-scale SNN processing systems, from both the modeling and implementation perspective, is to adopt a multi-core architecture design approach~\cite{Leite_etal22,Merolla_etal14a,Park_etal16,Davies_etal18,Moradi_etal18}.
In these systems each core consists of a neuro-synaptic array comprising digital or mixed-signal synapse and neuron soma circuits, and an asynchronous digital core interface.
The core interface is responsible for receiving input events and delivering them to the target synapses, and for transmitting the soma output spikes to target synapses and neurons, within the same core, or across multiple cores.

\begin{figure}
  \centering
  \includegraphics[width=1\linewidth]{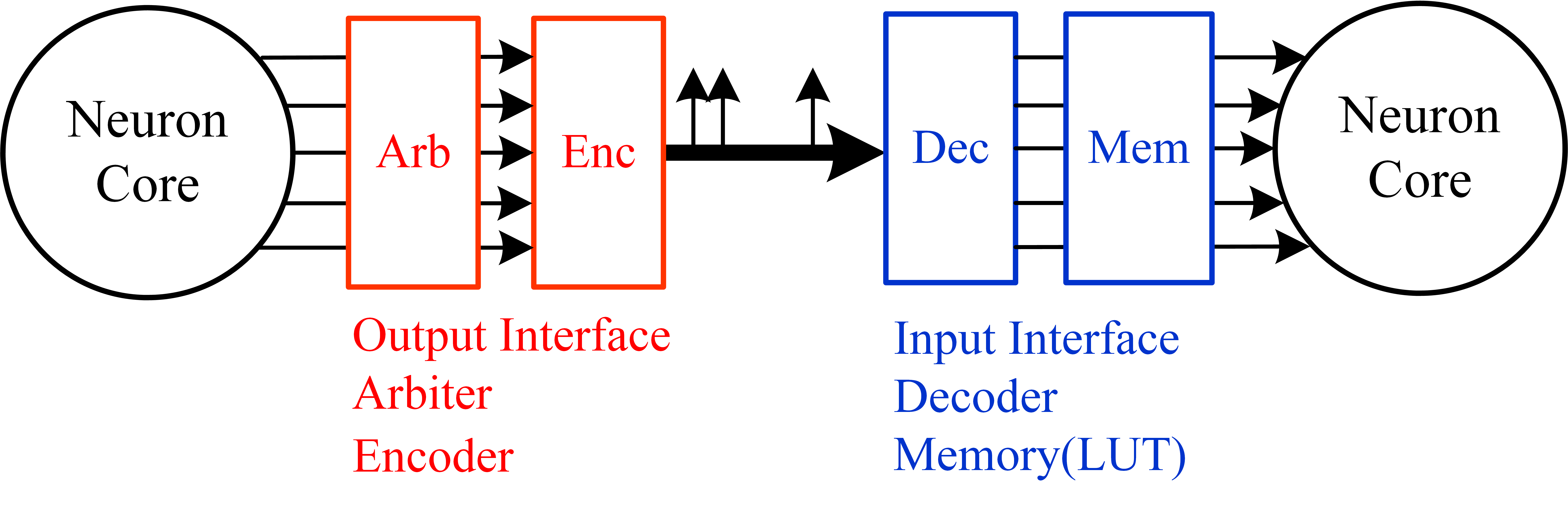}
  \caption{AER communication pipeline: each time a neuron spikes its address is encoded and transmitted on a shared bus using asynchronous circuits. Collisions (potential parallel spikes) are managed through asynchronous arbitration circuits, and neural network connectivity schemes are programmed via local memory Look Up Tables (LUT).}
  \label{fig1}
\end{figure}

The most common communication protocol use to transmit spikes from source neurons to destination ones in neuromorphic systems is based on the Address-Event Representation (AER)~\cite{Liu_etal14}. 
Figure~\ref{fig1} shows how the parallel output events from the neurons in a neuron core are encoded and time-multiplexed on a shared digital bus to provide support for inter-core and inter-chip communication.
An arbiter in the core output interface (``Arb'' in Fig.~\ref{fig1}) is used to manage potential collisions from multiple coincident neuron requests, and to grant access to the data bus to one neuron at a time.
The routing memory in the core input interface (``Mem'' in Fig.~\ref{fig1}) is used as Look Up Table (LUT) for storing and configuring the neural connections. A LUT can also be present in the output interface, depending on the different routing methods adopted.
An other type of memory that is commonly used in neuromorphic processors is the Content Addressable Memory (CAM), as its in-memory search operations can be instrumental for the network routing, when used directly in the synapse arrays of the neuromorphic cores. 

Despite recent improvements in ultra-low power neuron designs~\cite{Rubino_etal20} and high performance data packet switches~\cite{Bertozzi_etal20}, the optimization of the neuromorphic core interfaces remains a daunting task.
This is especially true for the arbitration architecture, when it includes the encoding pipeline (as is the case presented here), and the routing memory.
Both these elements represent the most important factors for the Power-Performance-Area (PPA) bottleneck of multi-core neuromorphic processors, as a function of neural network size.
For example, in the neuromorphic processors proposed in~\cite{Moradi_etal18}, the power consumption of the arbiter and routing memory takes up more than 80\% of the total power budget.

\subsection*{Contributions of this work}
In this work we substantially reduce the core interface hardware overhead for multi-core neuromorphic processors, by designing a novel asynchronous arbitration architecture in the core output interface and a new asynchronous CAM architecture in the core input interface. Specifically, the work presented:
\begin{itemize}
    \item provides a new arbitration mechanism based on a hierarchical arbiter tree (HAT) and its asynchronous encoding pipeline circuits.
    \item compares the new arbitration architecture to other existing arbitration architectures. 
    \item provides a new asynchronous CAM architecture based on CSCD, with feedback control and speculative sense. 
    \item presents custom-designed CAM circuits, with comparisons to conventional asynchronous CAM circuits.
\end{itemize}
We show how this arbitration architecture has improved performance, compared to previously proposed arbitration schemes, with up to 78.3\% lower latency figures and less area cost. To the best of our knowledge, no other asynchronous CAM architecture has been developed so far, which takes advantage of CSCD to perform robust search operations. The proposed CAM architecture achieves 40.4\% throughput increase and 46.7\% energy reduction with slight area increase compared to conventional asynchronous CAM arrays.

In the following section we discuss the background of arbiter and CAM circuits used in neuromorphic processors. In Section~\ref{sec:arbiter-tree} we present the new arbitration architecture. Section~\ref{sec:cam} presents the new CAM architecture, and in Section~\ref{sec:conclusion} we conclude the paper.

\section{Background}
\label{sec:background}

This section reviews the background on arbitration architecture and asynchronous CAM. It introduces the existing arbitration architectures and the asynchronous CAM, which forms the foundation of the new work of this paper.

\subsection{Arbitration schemes}
\textcite{Purohit_Manohar21} reviewed drawbacks and benefits of different arbitrating approaches. Arbitrating and encoding neuron's address based on a binary tree topology is suitable for applications with low event-rates and small neuron cluster size, because the request only needs to propagate through $\log_2(N)$ stages.
However, area cost and latency become worse when the neuron cluster size increases since the number of two-input arbiters increase linearly and every neuron's request has to propagate the whole arbiter tree.
The probability of grant overlapping also increases as the depth of arbiter tree increases.
The ``greedy tree'' represents an improvement to the original binary tree in situations where multiple input requests arrive within a very short time period~\cite{Boahen00}. But it suffers strict timing requirement which restricts the use for general applications~\cite{Boahen00}.
Both binary and greedy tree schemes have high power consumption, since each granted output of the arbiter drives $\log_2(N)$ address lines of the logarithmic encoder.

Another approach is to use a arbitrating mechanism with a ring-based topology.
This approach can quickly service a burst of localized events but becomes worse when sparse events are far apart in space, because the token has to travel for a long distance, for each input request, when requests are sparse.
\textcite{Purohit_Manohar21} propose a hierarchical token ring (HTR) method which can service sparse events like a binary tree and quickly scan through a section of the array like a linear token ring.
But it needs to change the number of processes in the rings and the number of levels of hierarchy to make the design tailored to different application scenarios, which is difficult to be implemented in a dynamic neuromorphic system since the neuron firing rates are dynamically changed. The high area cost of HTR also makes it hard to scale up the neuron core's size since the number of two-input arbiter also increases linearly as the number of neuron increases.

Here, we present a new arbitration mechanism based on multiple small arbiter tree and the circuits implementation of corresponding asynchronous encoding pipeline, which has lowest latency compared with all of other arbitration architectures when the events are sparse. During the burst event mode, HAT get similar performance with HTR and token-ring, but only needs $\log_2(N)$ two-input arbiters which makes it low area cost. Since HAT only use multiple small arbiter trees, which reduces the risk of grant overlapping in deep arbiter tree and makes the architecture more robust.

\subsection{CAM}
CAM cells have been widely used as a way to accelerate the search operation in large LUTs, due to their single-cycle parallel search  operation abilities~\cite{Lee_etal20}.
Neuromorphic processors usually use CAMs in addressable synapses to increase the flexibility of network mapping, especially when the network has  sparse connectivity~\cite{Moradi_etal18}.
Various CAM design approaches have been previously introduced. The CAM architecture based on the NOR-type CAM cell (reliable and fast)~\cite{Lee_etal20} and current-race match-line sense amplifier (MLSA) is widely used. This sensing scheme pre-charges the match-line (ML) low and evaluates the ML state by charging the ML with a current supplied by a current source. The benefits of this scheme over the precharge-high schemes are the simplicity of the threshold circuitry and the extra savings in search-line (SL) power due to the elimination of the SL precharge phase and also there is no charge-sharing problem~\cite{Pagiamtzis_Sheikholeslami06}.
In every CAM cell, in addition to a 6T-SRAM cell, there are three transistors for bit comparison. When the stored data and search data on the SL are same (in the \verb|MATCH| case), ML pull-down path (ML to GND) is disconnected then the ML can be charged until MLSA generates pulse as a input spike to the target neuron. On the other hand, when the stored-data and search data are opposite(in the \verb|MISMATCH| case), the ML pull-down path is formed and ML can't be charged. The \verb|Off| signal from the dummy CAM entry as shown in Fig.~\ref{fig6} is to terminate the current source in every MLSA, which is designed to be "always MATCH" with the worst case (assumed to be the last one to produce a \verb|MATCH| signal). Numerous switching of ML in \verb|MATCH| case and direct current flow in pull-down path in \verb|MISMATCH| case come at the cost of huge dynamic power consumption. Moreover, for event-driven neuromorphic processors, designing a robust asynchronous CAM architecture without sacrificing performance and energy efficiency is an another challenge, which is still a blank in field of asynchronous circuits.

\citeauthor*{Moradi_etal18}~\cite{Moradi_etal18} use the same CAM architecture described above as an asynchronous target memory with multiple tags. Each CAM entry (tag) represents the address of source neuron that the target neuron is subscribed to.
To minimize the area, this asynchronous CAM architecture is designed following a bundled-data style instead of (Quasi-delay-insensitive) QDI style.
The search operation of the asynchronous CAM architecture follows a standard four-phase handshaking protocol to communicate with the handshake(HS) block.
In order to guarantee the correct handshaking communication between the HS block and the CAM array, it is necessary to make two appropriate timing assumptions. The first one is that the presence of valid input data should be earlier than the request signal which is used to enable the searching operation. This is the common timing constraint in bundled-data design style which is not difficult to satisfy.
The second timing assumption is made when sending the acknowledge signal to HS block, to ensure that the search operation in the whole CAM array is completed. This assumption represents a key challenge for this asynchronous CAM architecture, because of the mismatch of current source circuits in MLSA and of the different numbers of \verb|MISMATCH| bits in the different CAM entries (which results in different ML wiring capacitance load).
These issues make it difficult to evaluate the time for completing the search operation and to make correct assumptions.
As shown in Fig.~\ref{6a}, a configurable delay line is used to leave enough timing margin for finishing the whole searching operation, which is a trade-off between performance and robustness.
To avoid the false negative error, high cycle time has to be the cost, which becomes the bandwidth bottleneck in multi-core neuromorphic processors. 

To address this problem we propose a novel asynchronous CAM architecture, which makes use of the CSCD technique to eliminate the second timing assumption.
CSCD exploits the fact that charging and discharging parasitic capacitance of internal nodes in digital circuits occur only when the signal is in transition to determine the working state of the circuit.
CSCD has already been used in asynchronous bundled-data pipeline circuits to take the advantages of the cost-efficient characteristics of bundled-data design without suffering the disadvantages of PVT-sensitive matching delay cells \cite{Huang_etal22}. The CSCD used in CAM architecture we propose is to detect the current flow change during the searching operation and act as an acknowledge signal generator. There are also two novel mechanisms in the new CAM architecture: feedback control and speculative sense to significantly reduce the power consumption in \verb|MATCH| and \verb|MISMATCH| cases respectively.

\section{Proposed hierarchical arbiter tree}
\label{sec:arbiter-tree}

In this section, the new arbitration mechanism and corresponding asynchronous encoding pipeline circuits are presented, followed by the experimental results and discussion.

\begin{figure} 
    \centering
  \subfloat[High level neuron cluster\label{2a}]{%
        \includegraphics[width=0.45\linewidth]{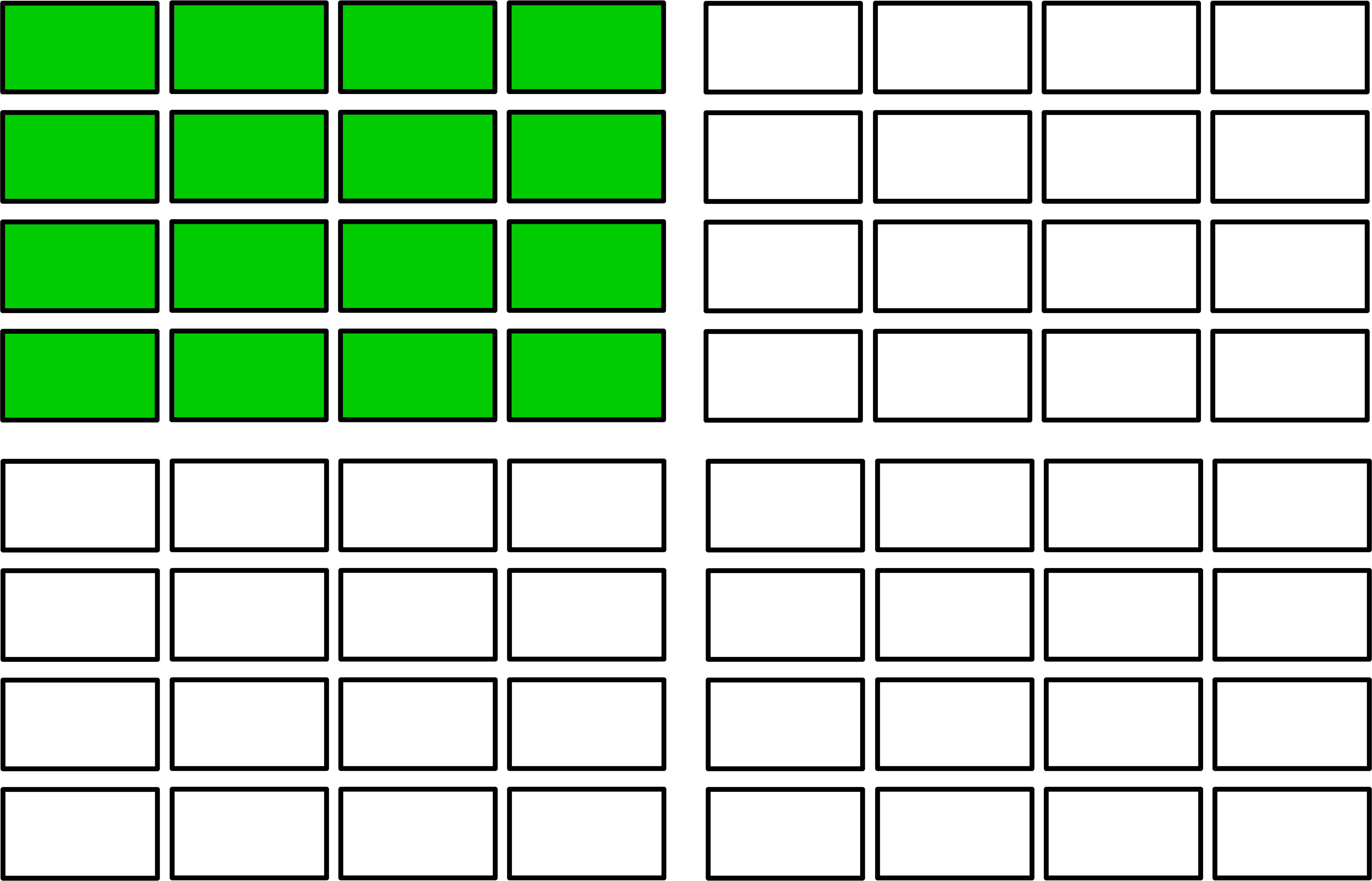}}
    \hfill
  \subfloat[Medium level neuron cluster\label{2b}]{%
        \includegraphics[width=0.45\linewidth]{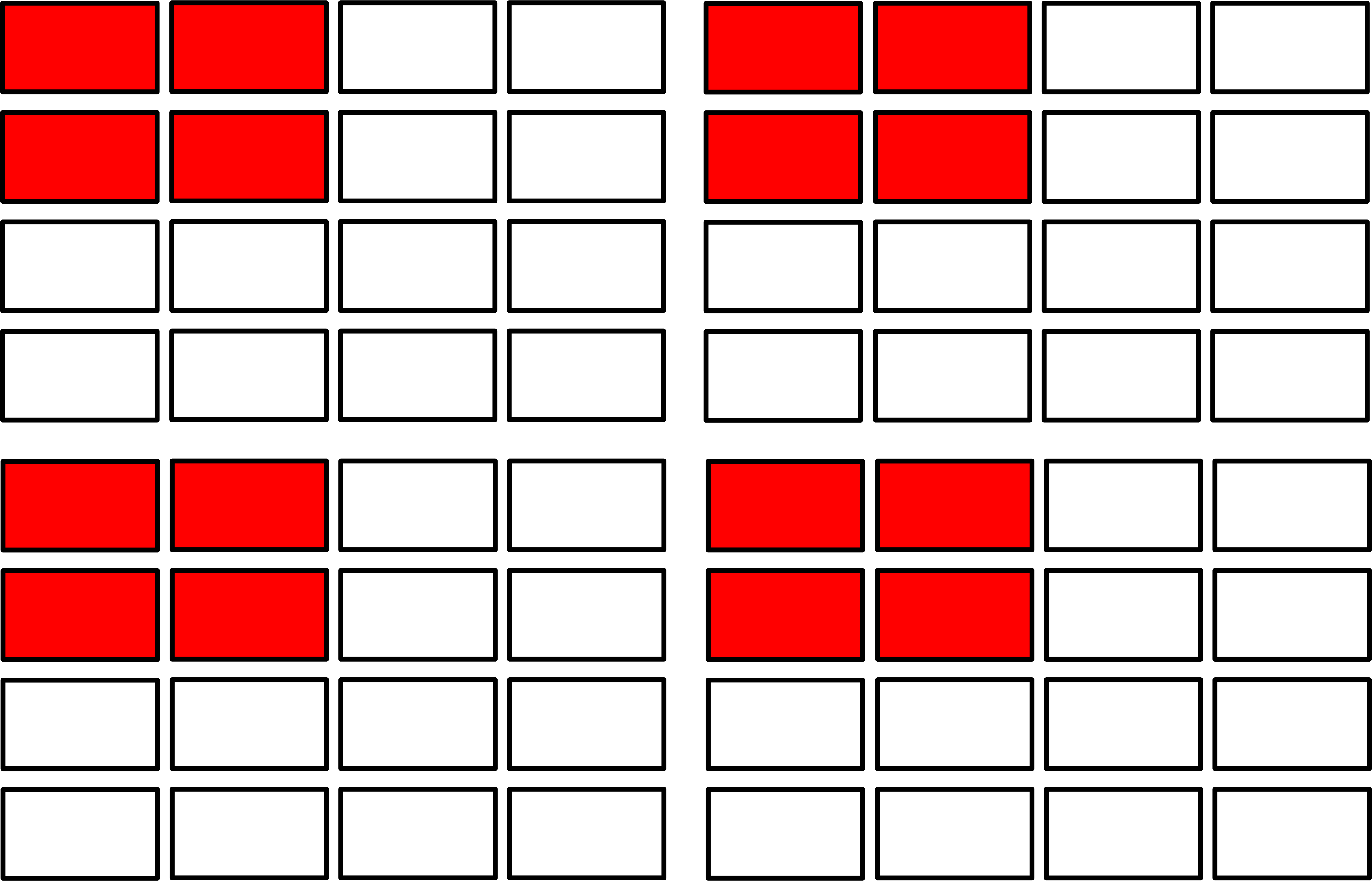}}
    \\
  \subfloat[Low level neuron cluster\label{2c}]{%
        \includegraphics[width=0.45\linewidth]{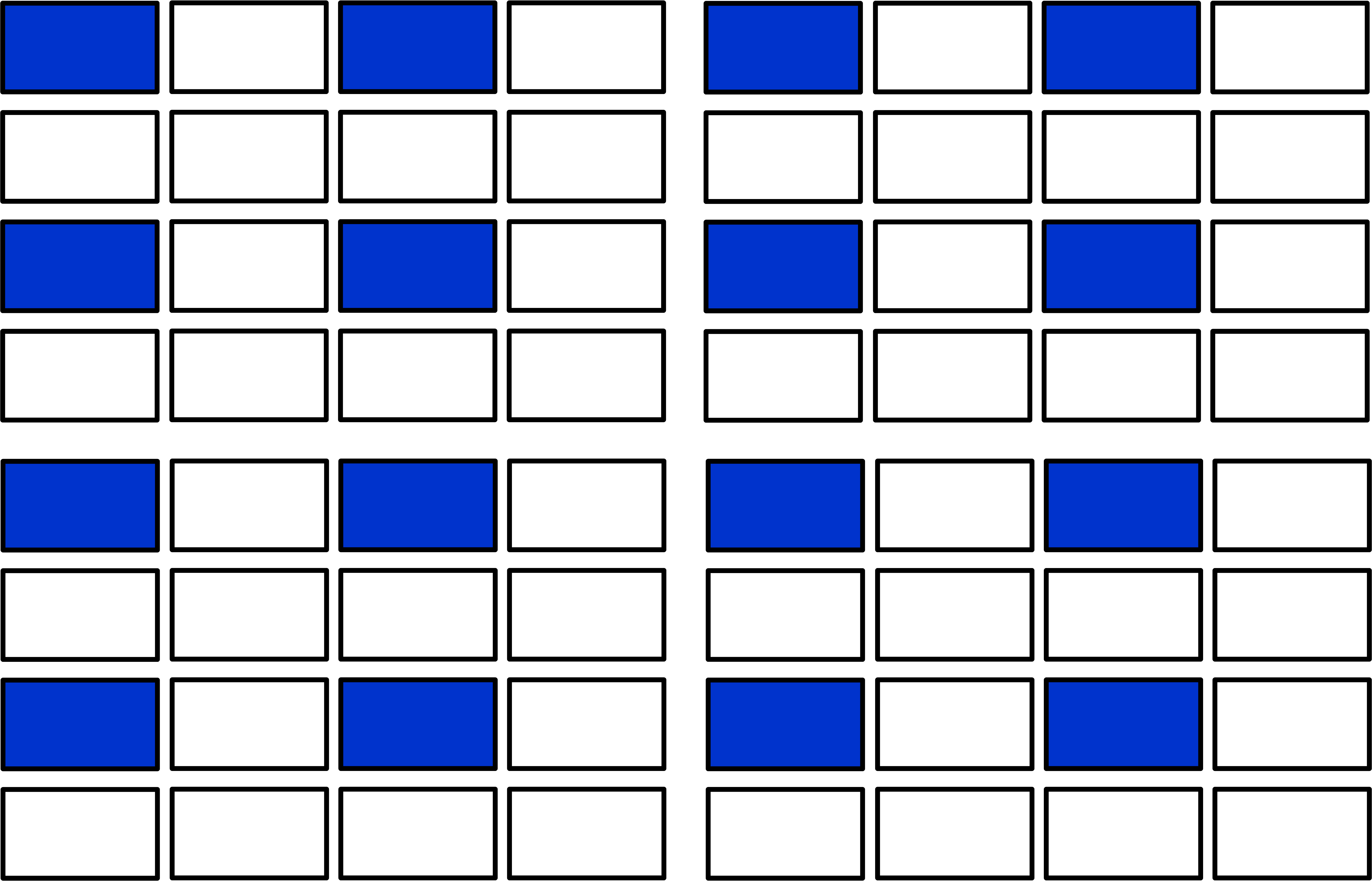}}
  \caption{Hierarchical arbitration mechanism schemes.}
  \label{fig2} 
\end{figure}

\subsection{Arbitration Mechanism}
Figure~\ref{fig2} shows an example of 64 neurons encoded by a hierarchical arbitration mechanism, which needs a 6four-inputs deep arbiter tree to encode 64 neurons using 6 bits if there is no any hierarchical arbitration. Based on HAT method, arbitration and encoding can be done for every 2bits. As shown in Fig.~\ref{2a}, the cluster with 16 neurons (highlight in green) share the pin \verb|Req[0]| and \verb|Grant[0]| in high level arbiter "ArbiterH" in Fig.~\ref{fig3}, which is usually implemented by pull-down transistors and pull-up circuits to reduce area cost instead of using OR gate tree~\cite{Boahen00,Moradi_etal18}. The neurons highlight in Fig.~\ref{2b} and Fig.~\ref{2c}, which share the pin \verb|Req[0]| \verb|Grant[0]| of medium level arbiter "ArbiterM" and the pin \verb|Req[0]| \verb|Grant[0]| of  low level arbiter "ArbiterL" respectively. The arbitration starts from high level arbitration. Only when the arbiter gives the grant to one of the four neuron clusters, the active neurons in that cluster can send the requests to the medium level arbiter. The operation relationship between the medium level arbiter and the low level arbiter is the same as that. The arbiter will not give grant to another cluster until all of active neurons in the current cluster have been encoded.

\subsection{Asynchronous encoding pipeline circuits}
\begin{figure}
  \centering
  \includegraphics[width=1\linewidth]{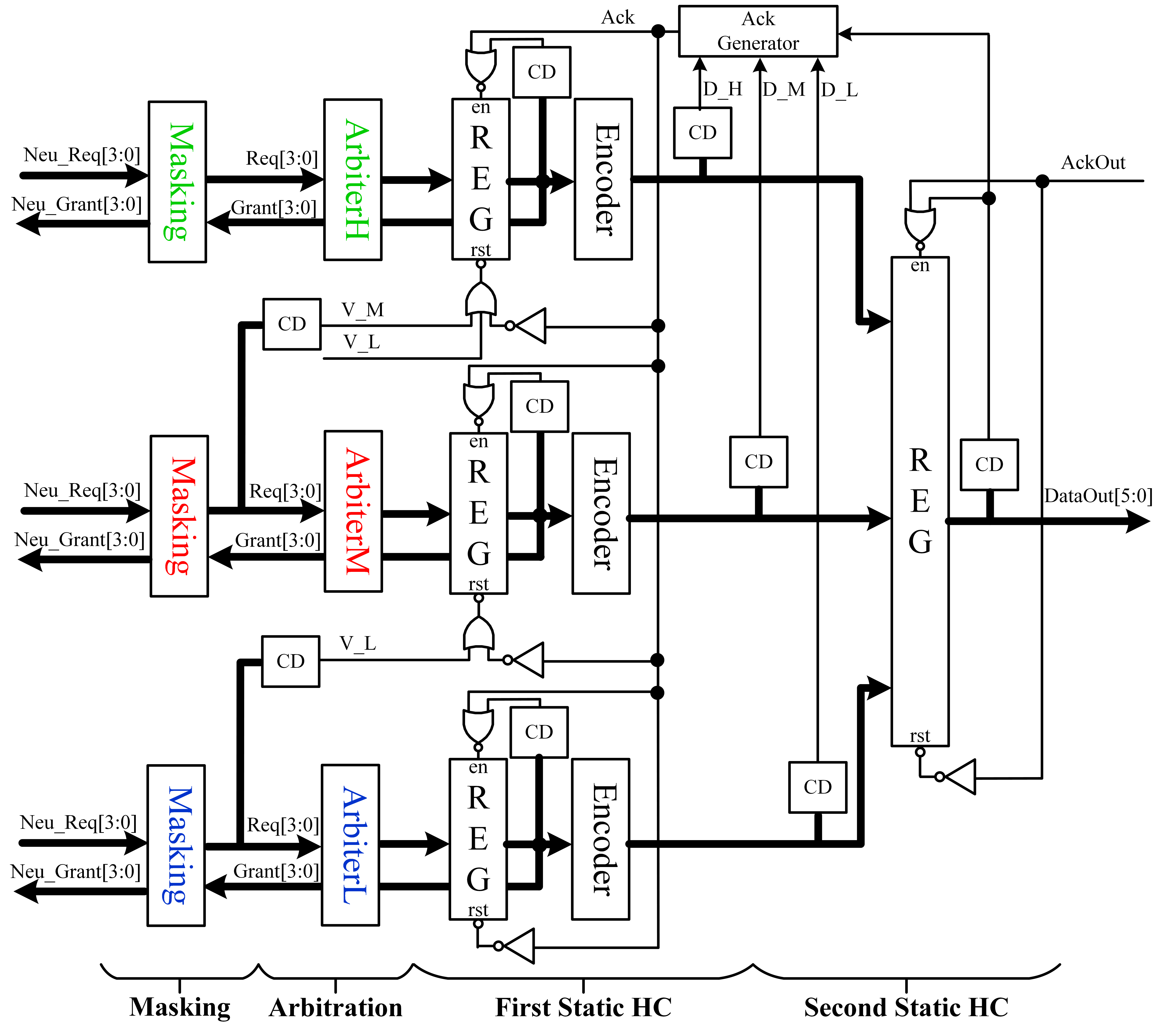}
\caption{Asynchronous pipeline circuits.}
\label{fig3}
\end{figure}

\begin{figure} 
    \centering
  \subfloat[\label{4a}]{%
        \includegraphics{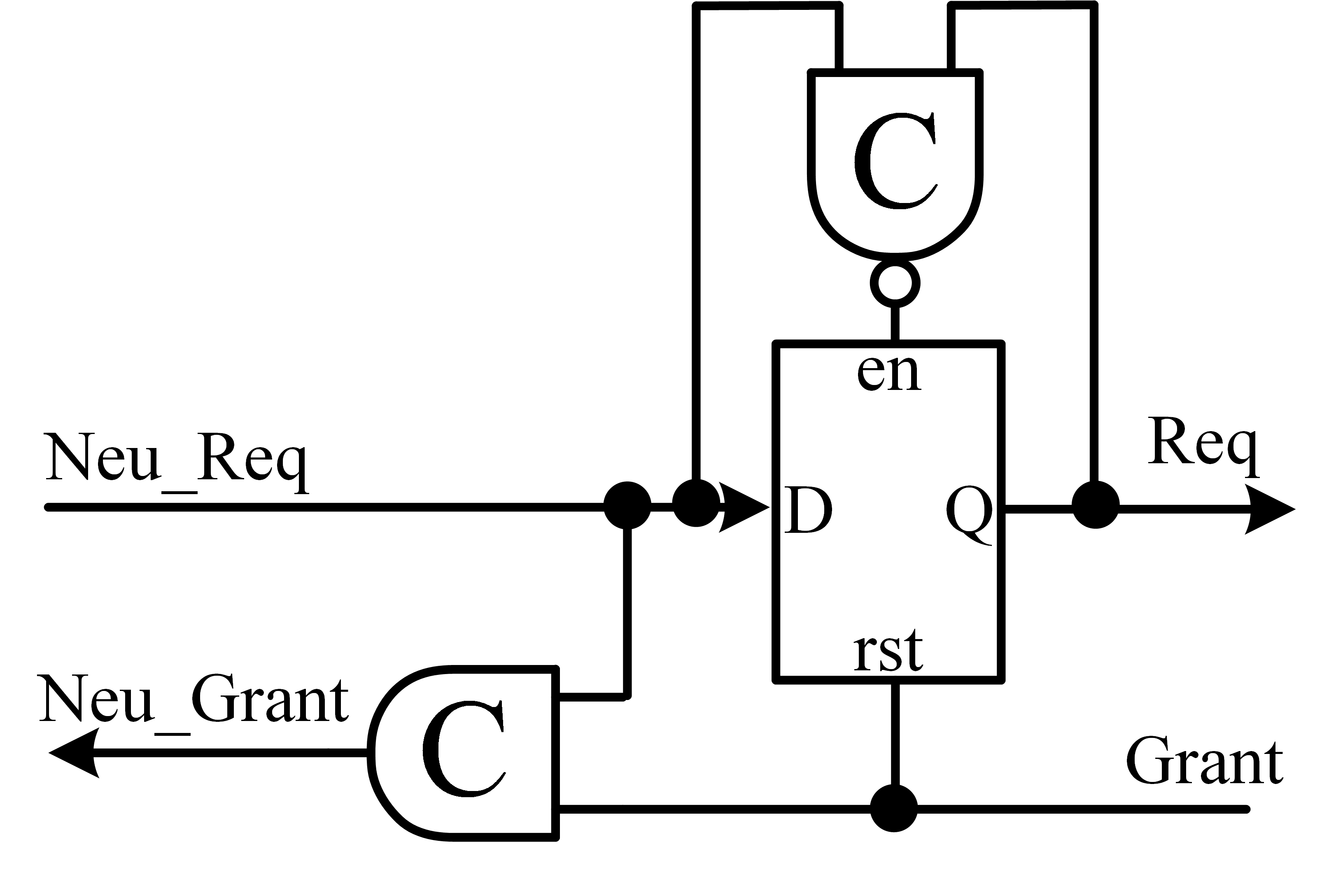}}
    \hfill
  \subfloat[\label{4b}]{%
        \includegraphics{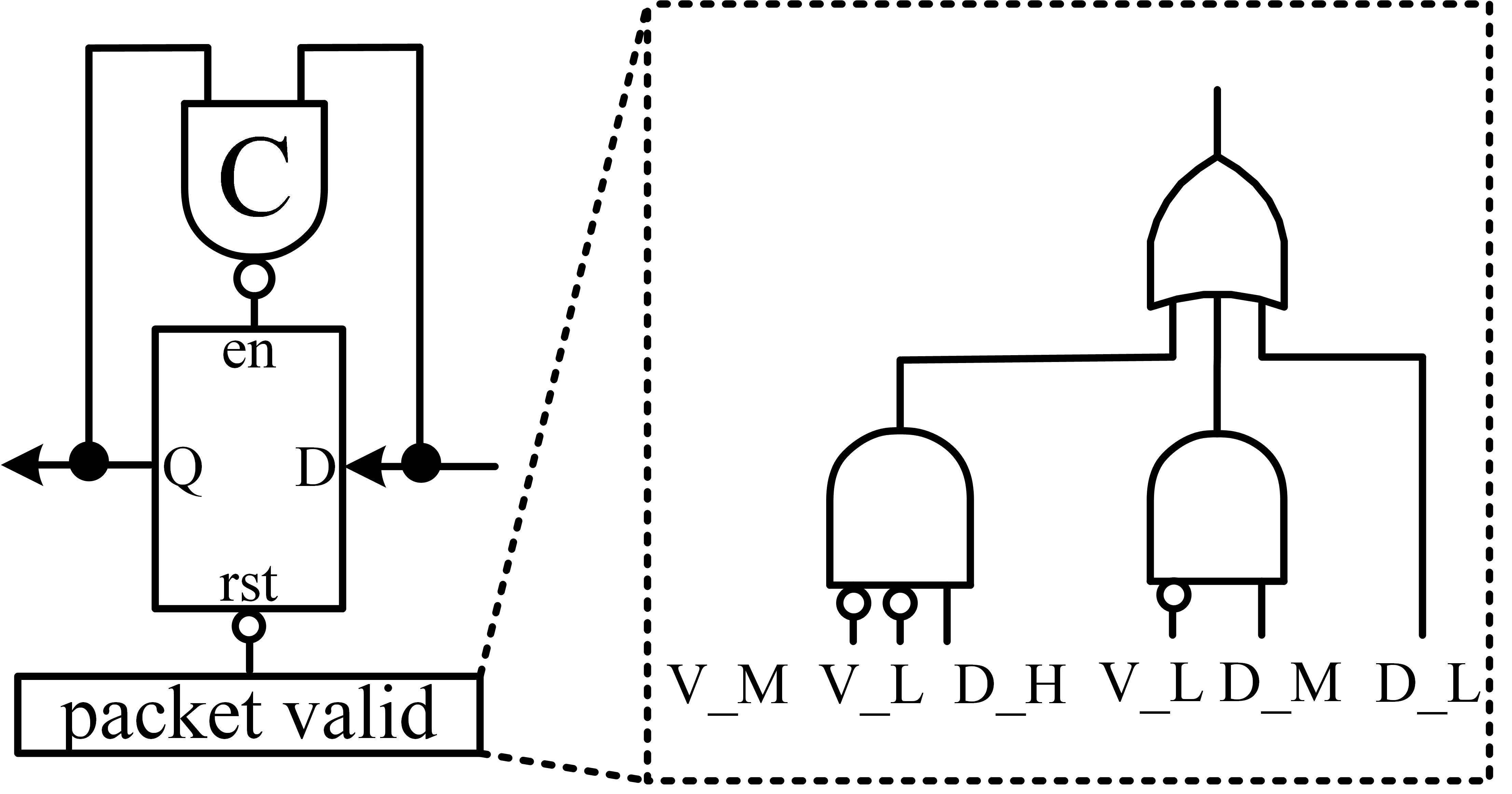}}
  \caption{(a) Masking gate; (b) Ack generator.}
  \label{fig4} 
\end{figure}

Inspired by high-capacity dynamic pipeline using static logic (static HC) in~\cite{Jiang_Nowick17}, HAT is implemented as shown in Fig.~\ref{fig3}. Three levels hierarchical arbitration is shown as an example and every hierarchy level uses a low cost four-input arbiter trees. The output of this asynchronous pipeline circuits is used as the LUT pointer index to get routing data packet or directly sent to the Network-on-Chip (NoC). The four-phase QDI circuits is used here because it's more compatible with the neuron handshake circuits and robust. The working flow is divided by four stages. The first stage is "masking stage", as shown in Fig.~\ref{4a}, which adopts the static logic to decouple the long-term handshaking protocol in neuron handshake circuits, from the rapid handshaking and release in the arbiter. Compared with~\cite{Jiang_Nowick17}, a C-element is added to make the handshaking strictly follow four-phase handshaking and more robust. The complete detection (CD) block after masking gate is to detect if there are still active requests from low level or medium level, if so, the circuits can't release the medium level grant or high level grant respectively. This means the architecture doesn't need to encode higher level bits every time when it handles lower level data packet handshaking, which is more energy efficient. 

The second stage is "arbitration stage", which provides a one-hot output to the third stage "first static HC pipeline". The one-hot output from the third stage is sent back as grant signals to reset arbiter's input, and at the same time it acts as input of the QDI encoder. The CD block after "first static HC pipeline" is used to test if the output is valid. To avoid grant overlapping problem of arbiter, CD block here is consisted of XOR gates instead of OR gates. The output data of QDI encoder will be sent to the last stage "second static HC pipeline", which merges the encoded data from three levels as a complete data packet including 6 bits that represents neuron address. The CD block after the "second static HC pipeline" is used to evaluate the complete data packet and acknowledge the previous stage. 

The outputs of three CD blocks after the QDI encoder of the "first static HC pipeline" are used to evaluate if the data of the corresponding level is still valid and reset ack generator if not, which deasserts \verb|Ack| signal to enable the "first static HC pipeline" and complete an entire cycle. As shown in Fig.~\ref{4b}, the CD block in the low level resets ack generator whenever a complete 6 bits data packet is captured by the "second static HC pipeline". The medium level CD block and high level CD block can only reset ack generator successfully when there is no valid neuron request in lower level.

\subsection{Timing Analysis}
The proposed arbitration architecture involves three pipeline-related timing constraints, the first two of which are directly transformed from those in the original static HC pipeline~\cite{Jiang_Nowick17}. \emph{(i)} The first timing constraint is the hold timing of the pipeline register. The closing of latch should be earlier than data reset on the input channel, which can be simply satisfied since there is a round-trip communication to reset the input data.

\emph{(ii)} The second timing constraint is that the current register should be re-opened later than the input data reset. This timing constraint is also easy to satisfy because the re-open operation needs a complete four-phase handshake.

\emph{(iii)} The third timing constraint is related to the HAT mechanism, which is in the Ack generator. The \verb|V_M| and \verb|V_L| should be changed faster than the \verb|D_H| \verb|D_M| and \verb|D_L|, otherwise the Ack generator will be reset wrongly and re-open the "first static pipeline stage" too early. In practice, this timing constraint is simple to satisfy, since the CD block after masking registers provides output whenever the input data is valid, however the CD block after the QDI encoder needs to wait for the valid data going through the arbiter, the first static HC pipeline and the encoder.

\subsection{Experimental results and discussion}
Table \ref{tab1} \ref{tab2} and \ref{tab3} shows the comparison of theoretical calculation results and pre-layout results between HAT and other arbitration architectures. All arbitration architecture designs are mapped using a 22FDX FDSOI standard cell library. Analog mutual exclusion elements (mutexes) in the two-input arbiters are implemented with standard-cell equivalent version~\cite{Ghiribaldi_etal13}. Gate size is decided by SPICE simulations and use \verb|set_dont_touch| command during synthesis to avoid any optimization on it. The asynchronous sequential C-elements are implemented using combinational gates with feedback. We care more about latency than throughput is because SNNs is more sensitive to the temporal information. On the other hand, the neuron handshake circuit usually has several stages of pipeline buffers and the time interval of two neuron spikes is longer than the arbitration encoding time, which will relax the requirement on throughput. All latency results are in typical operating conditions. The latency of greedy tree in burst mode is not considered here because it highly depends on the response time of the neuron, which is the same reason as it in~\cite{Purohit_Manohar21}. The synthesis tool flow is similar with~\cite{Miorandi_etal17}. We use the generic GTECH Synopsys library to implement the very low-level but technology-independent specification, which can help us has full control over the gate-level logic function. During the synthesis, only gate sizing and buffer insertion are allowed. The \verb |set_max_delay| command is applied to all of the timing paths in order to get high performance. The clock and reset paths have higher weight of delay constraint during technology mapping, which is to avoid violations of minimum pulse width and hold time . 

Two different cases (sparse event and full frame burst event) similar with~\cite{Purohit_Manohar21} are considered. A random event request from N neurons is selected and the latency is measured from neuron request to output request. In the asynchronous pipeline circuits we proposed, the output request is the output signal of last pipeline's CD block, which represents the complete neuron address data packet is valid. The average latency is measured for N neurons. For full frame burst events, all neurons fire in the short time window, which is the starting point of latency measurement, then the latency is measured between staring point and last output request signal. For the theoretical calculation results of latency, we assume the latency in moving the handshake signal between two stages is small compared to the latency of handling the events and normalized by two-input arbiter's latency. 

For the theoretical calculation results of area, the area cost is evaluated by the number of two-input arbiter since most of the area will be occupied by two-input arbiters as neuron number increases. The latency and area for each arbitration architecture are estimated based on the expression for 64 and 256 neurons. For HAT, every hierarchical level has a four-input arbiter which is the same architecture as~\cite{Ghiribaldi_etal13}, then 64 neurons and 256 neurons need three hierarchical levels and four hierarchical levels respectively. For HTR, two-level hierarchical token-ring is used and every level has $\sqrt{N}$ leaf nodes, which is the same as the experiment in~\cite{Purohit_Manohar21}.

\begin{figure} 
    \centering
  \subfloat[Average latency with sparse events.\label{5a}]{%
        \includegraphics[width=0.475\linewidth]{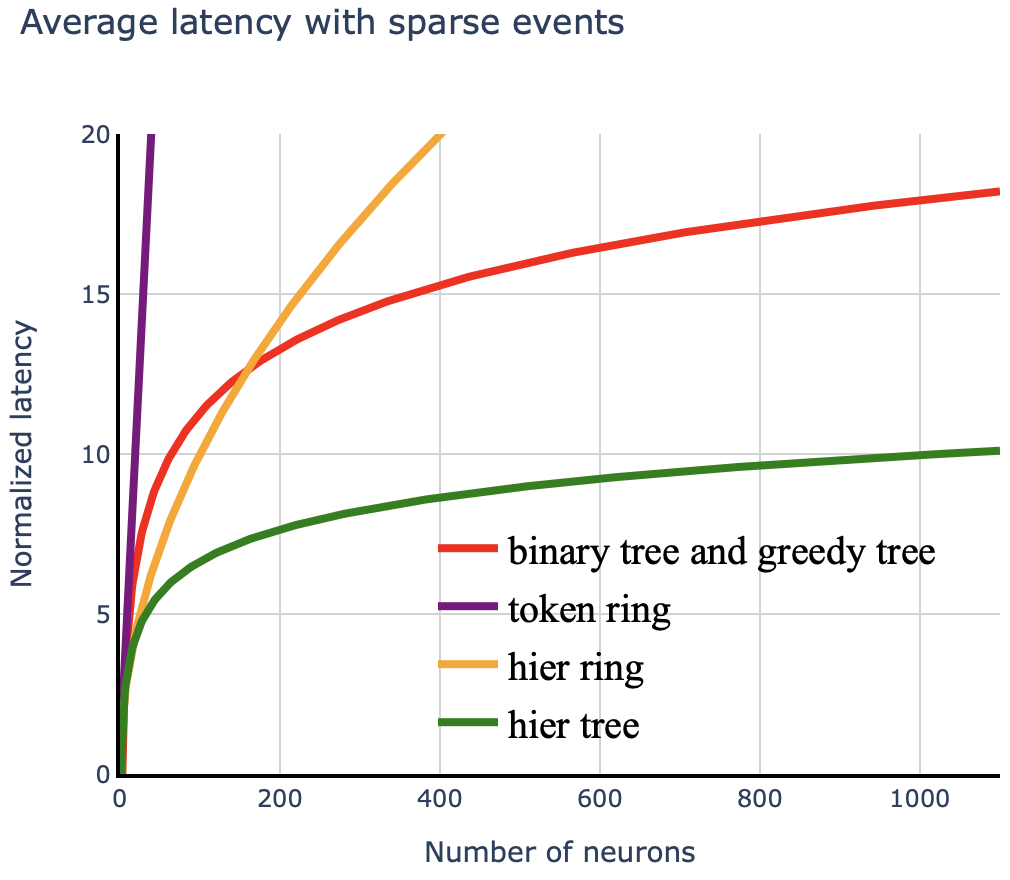}}
    \hfill
  \subfloat[Average latency with burst events.\label{5b}]{%
        \includegraphics[width=0.475\linewidth]{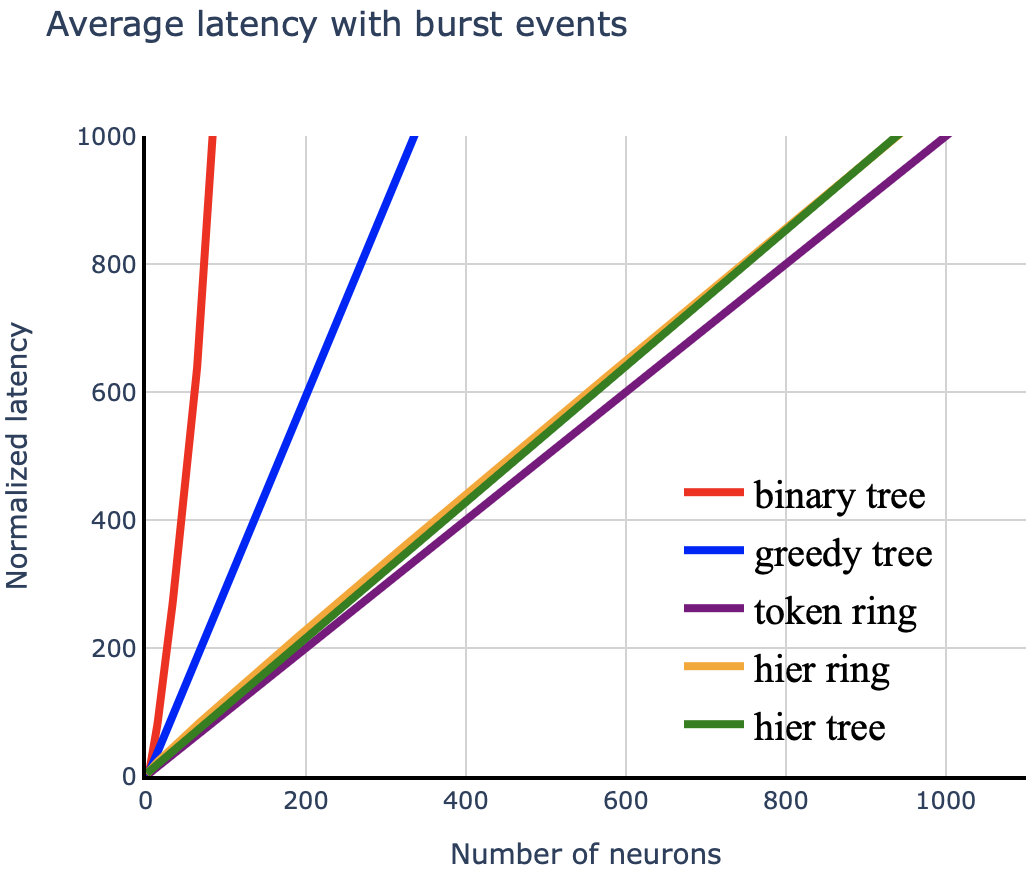}}
  \caption{The scalability of latency.}
  \label{fig5} 
\end{figure}

It is shown that HAT performs best when the arbitration architecture has to support both sparse event mode and burst event mode. As shown the pre-layout results in Table \ref{tab1}, HAT offers significant improvements in latency when the event is sparse, which is the general case for neuromorphic processors. Thanks to static HC pipeline used in asynchronous encoding pipeline circuits, HAT can still achieve low latency even if the event needs to go through several pipeline stages. Fig.~\ref{5a} shows the scalability of arbitration latency as the cluster size increases in sparse events mode. It can be seen HAT has the lowest latency. For burst event mode, while HAT is slightly slower than token ring as shown in Table \ref{tab2}, token ring has much higher latency in sparse event case. It can also be observed in Fig.~\ref{5b},  in burst event mode HAT has similar latency compared with token-ring and HTR as the number of neuron increases and all of these three methods have significantly lower latency than binary and greedy tree.
More importantly, HAT uses much less two-input arbiters as the number of neuron increases as presented in Table \ref{tab3}, since the number of two-input arbiters used in other method increases linearly as the neuron core size increases. We normalized the area cost of the whole arbitration architecture by the area of two-input arbiter cell as shown in Table \ref{tab3}. The area of HAT doesn't scale up as the theoretical calculation, which is because that every neuron handshake circuits has $\log_2(N)$ pull down transistors and every hierarchical cluster has a pull up circuits for sharing the arbiter. These circuits results in some extra area cost. Using $\log_2(N)$ small arbiter trees in HAT not only reduces area cost and potentially increases energy efficiency, but also reduce the probability of grant overlapping existing in deep arbiter trees.
 
\begin{table}
\caption{Theoretical and simulated average latency with sparse events}
\begin{center}
\begin{tabular}{|c|c|c|c|}
\hline
\textbf &\multicolumn{3}{|c|}{\textbf{Average latency with sparse events}} \\
\cline{2-4} 
\textbf & \textbf{\textit{Latency}}& \textbf{\textit{N=64}}& \textbf{\textit{N=256}} \\
\hline
Binary tree & $2*(\log_2 N-1)$ & 10 (1.7\,$ns$) & 14 (2.1\,$ns$) \\
\hline
Greedy tree & $2*(\log_2 N-1)$ & 10 (1.8\,$ns$) & 14 (2.3\,$ns$) \\
\hline
Token-ring & $(N+1)/2$ & 32.5 (25.3\,$ns$) & 128.5 (102.7\,$ns$)  \\
\hline
Hier-ring & $\sqrt{N}$ & 8 (5.7\,$ns$) & 16 (9.2\,$ns$) \\
\hline
Hier-tree & $\log_2 N$ & 6 (1.7\,$ns$) & 8 (2.0\,$ns$)\\
\hline
\end{tabular}
\label{tab1}
\end{center}
\end{table}

\begin{table}
\caption{Theoretical and simulated average latency with burst events}
\begin{center}
\begin{tabular}{|c|c|c|c|}
\hline
\textbf &\multicolumn{3}{|c|}{\textbf{Average latency with burst events}} \\
\cline{2-4} 
\textbf & \textbf{\textit{Latency}}& \textbf{\textit{N=64}}& \textbf{\textit{N=256}} \\
\hline
Binary tree & $2N*(\log_2 N-1)$ & 640 (83.7\,$ns$) & 3584 (436.9\,$ns$) \\
\hline
Greedy tree & $3N-6$ & 186 & 762 \\
\hline
Token-ring & $N$ & 64 (40.5\,$ns$) & 256 (178.4\,$ns$) \\
\hline
Hier-ring & $N+2\sqrt{N}$ & 80 (48.9\,$ns$) & 288 (192.9\,$ns$) \\
\hline
Hier-tree & $\frac{17}{16}N+3$ & 71 (47.2\,$ns$) & 275 (194.4\,$ns$) \\
\hline
\end{tabular}
\label{tab2}
\end{center}
\end{table}

\begin{table}
\caption{Normalized area cost}
\begin{center}
\begin{tabular}{|c|c|c|c|}
\hline
\textbf &\multicolumn{3}{|c|}{\textbf{Normalized area cost}} \\
\cline{2-4} 
\textbf & \textbf{\textit{Number of two-input arbiter}}& \textbf{\textit{N=64}}& \textbf{\textit{N=256}} \\
\hline
Binary tree & $N-1$ & 63 (72.3) & 255 (277.4) \\
\hline
Greedy tree & $N-1$ & 63 (83.4) & 255 (286.7) \\
\hline
Token-ring & $N$ & 64 (79.1) & 256 (272.5)  \\
\hline
Hier-ring & $N+2\sqrt{N}$ & 80 (89.2) & 288 (296.3) \\
\hline
Hier-tree & $3\log_4 N$ & 9 (59.4) & 12 (192.4)\\
\hline
\end{tabular}
\label{tab3}
\end{center}
\end{table}

\section{Proposed CAM architecture}
\label{sec:cam}

Given the issues of low performance and low energy efficiency in the conventional asynchronous CAM architecture, this section introduces a new CAM architecture with CSCD, and also the mechanisms of feedback control and speculative sense in MLSA. 

\subsection{CAM Architecture with CSCD}
\begin{figure} 
    \centering
  \subfloat[\label{6a}]{%
        \includegraphics{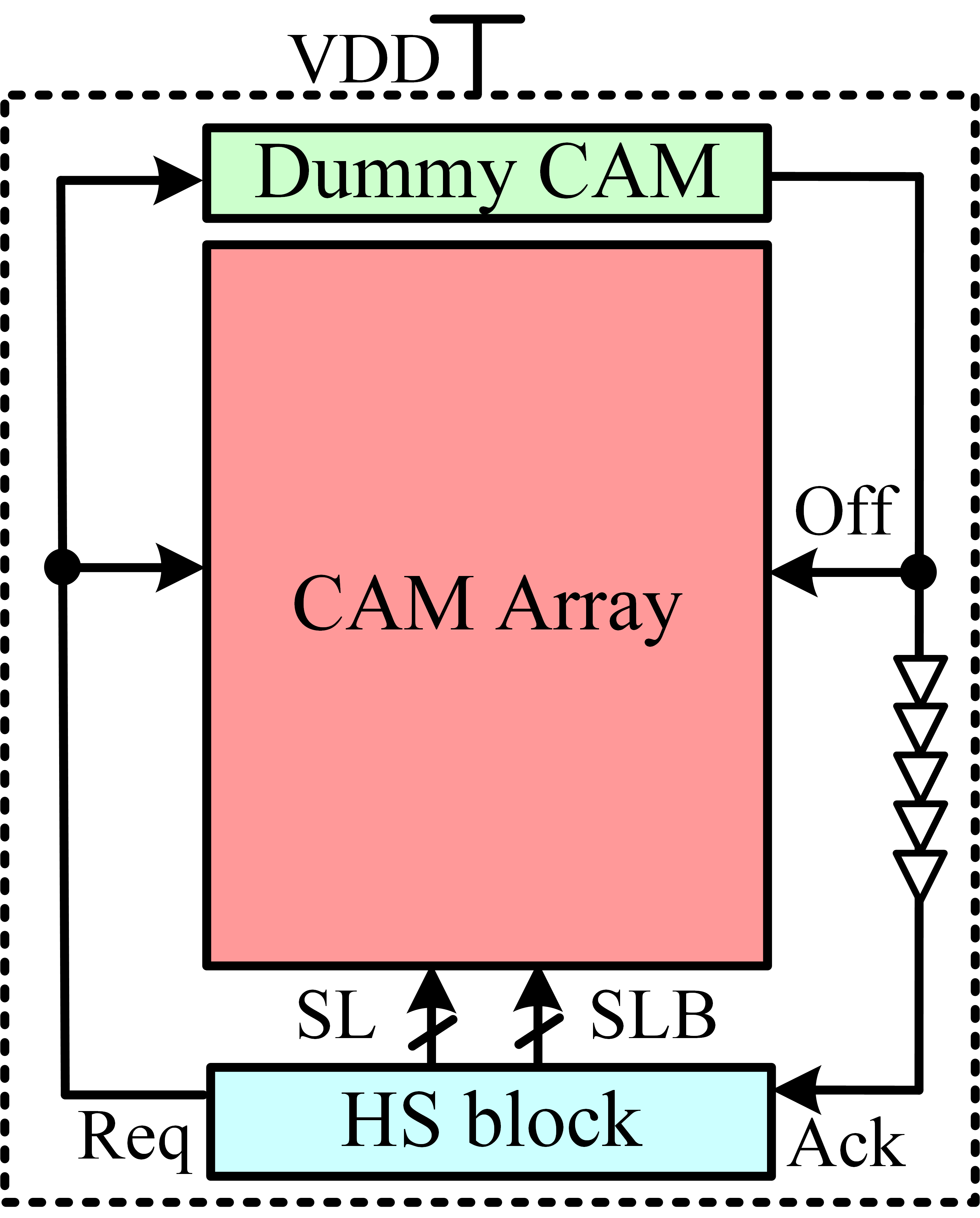}}
    \hfill
  \subfloat[\label{6b}]{%
        \includegraphics{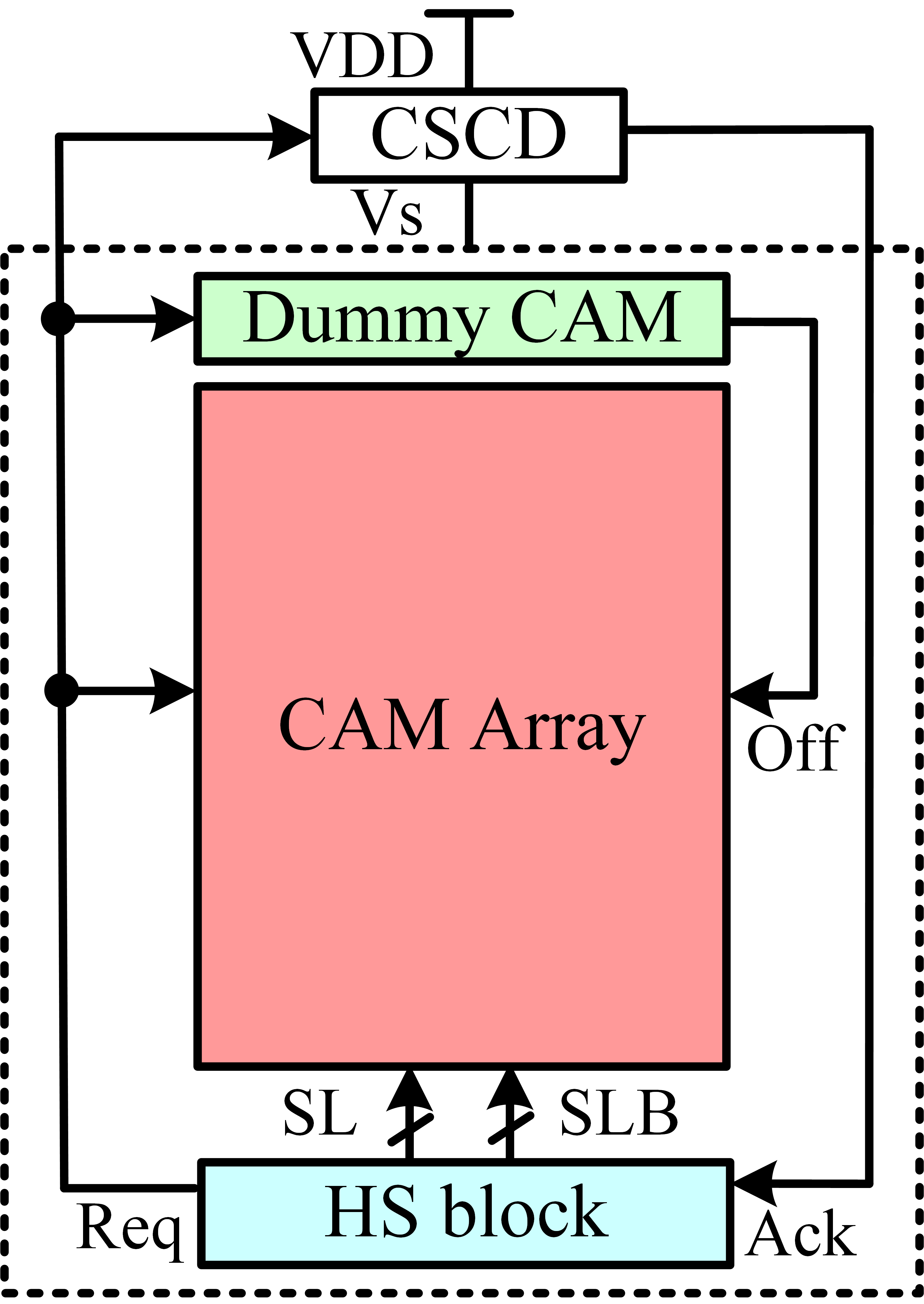}}
  \caption{(a) Conventional asynchronous CAM architecture; (b) Asynchronous CAM architecture with CSCD.}
  \label{fig6} 
\end{figure}

Fig.~\ref{fig6} shows the difference between the conventional asynchronous CAM architecture and the new CAM architecture we propose. In conventional asynchronous CAM architecture, the request signal from handshake(HS) block is sent to CAM array and the dummy CAM entry in parallel. Here we assume the request and acknowledge signals follow a four-phase handshake protocol. The always on dummy CAM entry provides the \verb|MATCH| signal whenever it receives the request signal, which is used as the \verb|Off| signal to terminate charging the ML in all of the CAM entries and also sent back to the HS block as an acknowledge signal after a configurable delay line. The delay here results in high cycle time, which is also a trade-off between performance and robustness. To solve this key challenge, we propose the CAM architecture with CSCD block as presented in Fig.~\ref{6b}. The first concept of a CSCD sensor was published by~\cite{Dean_etal94}. The CSCD sensor is inserted between the logic function unit and power supply to detect current flow. The sensor produces a low output when no current flowing through the logic (i.e., the logic is not working), and produces a high output when the combinational logic transitions. Here we use the CSCD block to evaluate the current flowing through the CAM Array during search operation. 

\begin{figure}
\centerline{\includegraphics[width=1\linewidth]{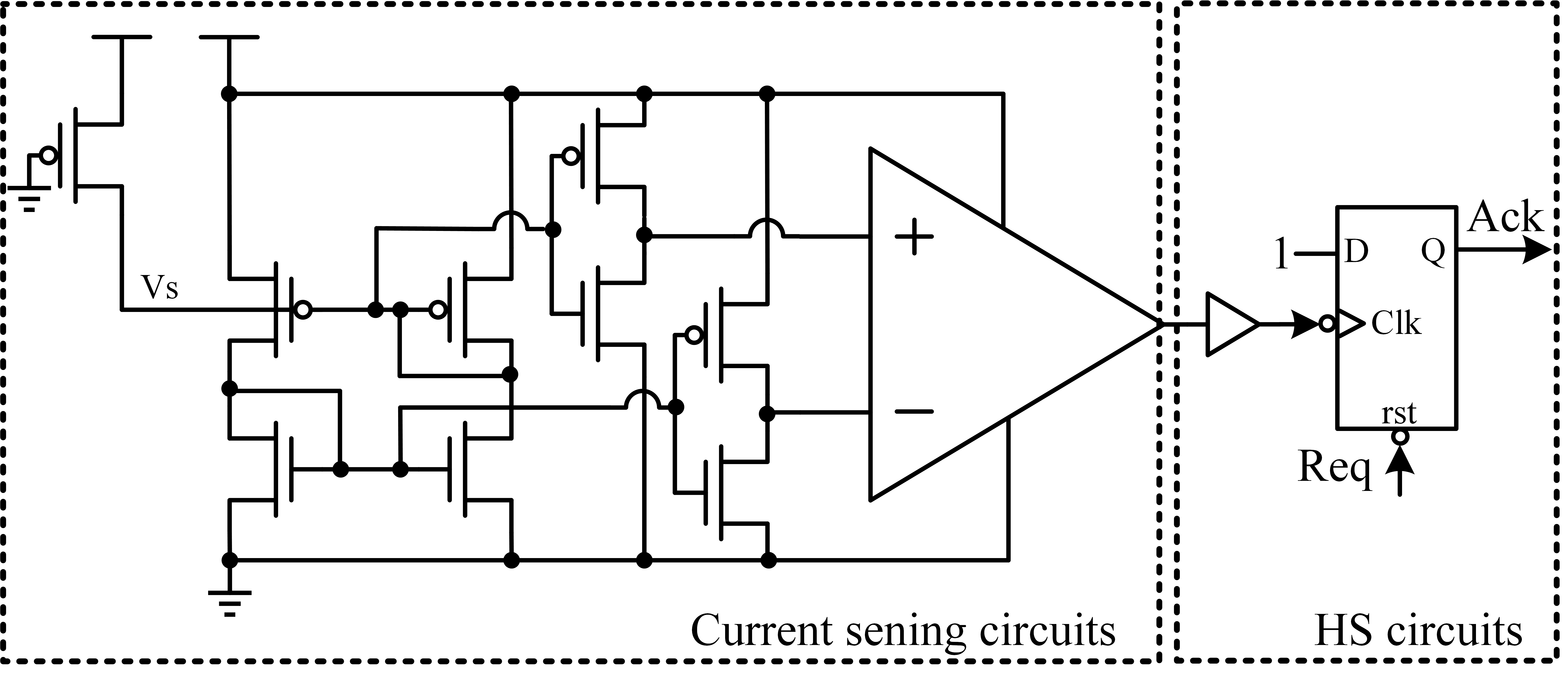}}
  \caption{CSCD block.}
\label{fig7}
\end{figure}

As shown in Fig.~\ref{fig7}, CSCD block includes the current sensing circuits and HS circuits which does four-phase handshake with HS block of the CAM architecture. The complete working flow of this CAM architecture is that (a)HS block provides a request signal after the data is valid on the SL and SLB to enable the searching operation of the CAM array, and deasserts the reset signal of register in the CSCD block. (b)Current sensing block evaluates the current flowing through the CAM array and generates the rising edge when the CAM array is doing searching operation. (c)After the \verb|Off| signal from the dummy CAM entry terminates charging ML, the falling edge from current sensing circuits triggers the register to provide the acknowledge signal. (d)The HS block deasserts the request signal after it receives the acknowledge signal from CSCD, which is used to precharge ML in CAM array to GND and also reset acknowledge signal in CSCD block to finish the whole four-phase handshake. The current sensing circuits is basically similar with it in~\cite{Huang_etal22}. Usually during the operation of precharging ML to GND, the current change can't make current sensing circuits generate a pulse since the duration of current change is too short for the high speed amplifier to detect. The same is true during CAM writing operation.

\subsection{Feedback control and speculative sense}
\begin{figure}
\centerline{\includegraphics[width=1\linewidth]{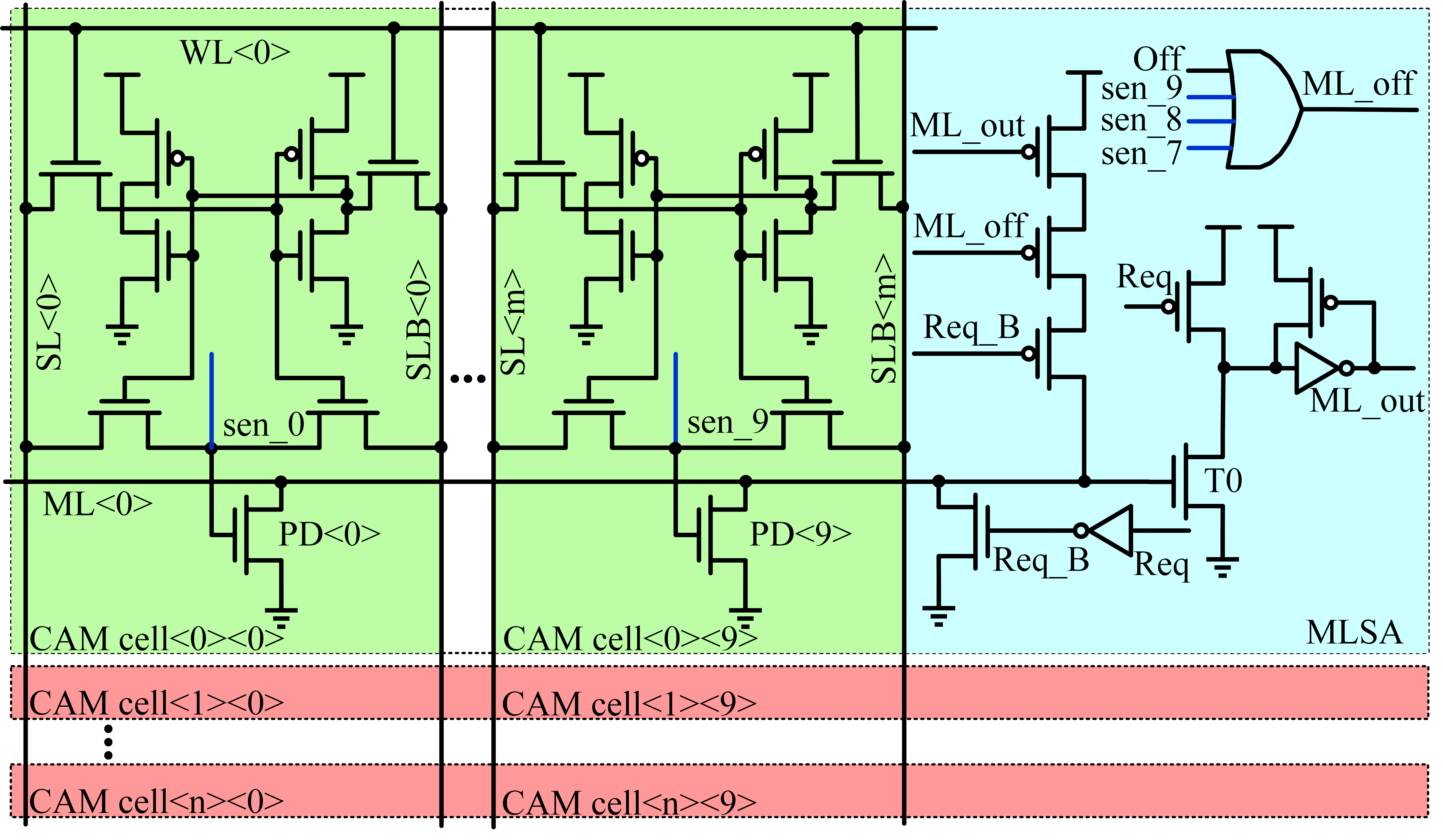}}
\caption{CAM Array with feedback control and speculative sense.}
\label{fig8}
\end{figure}

In order to reduce the dynamic power during searching operation. We propose the mechanisms of feedback control and speculative sense in MLSA. Fig.~\ref{fig8} shows the CAM array based on NOR-type CAM cell and current-race MLSA with feedback control and speculative sense, which has \verb|n| CAM entries and 10 CAM cells in each CAM entry. The feedback control and speculative sense can reduce power in \verb|MATCH| and \verb|MISMATCH| respectively. In \verb|MATCH| case, the current source in MLSA charges the ML until the voltage is higher than the threshold of the transistor \verb|T0|, which makes MLSA generate a high output. Here we use the output signal as a feedback signal to close current source. Feedback control mechanism makes the MLSA do adaptive sensing and quickly close charging by itself when the CAM entry is \verb|MATCH|, so there is no need to wait for the \verb|Off| signal from dummy CAM entry, which usually reduces around 40\% voltage swing on ML in \verb|MATCH| case. 

In \verb|MISMATCH| case, direct current flowing through ML pull-down path causes a significant dynamic power consumption. The basic idea here is still trying to close the current source as soon as possible. Since the gate voltage of the tail transistor \verb|PD| in every CAM cell changes quickly after the data on the SL and SLB is valid, which can be used to do the early detection whether the data bit in CAM cell is \verb|MATCH| or \verb|MISMATCH|. We add extra one pin \verb|sen_n| in every CAM cell as shown in Fig.~\ref{fig8} to sense the matching status before the searching operation(the \verb|Req| arrives). The signal from the sense node will go through the OR gate in the MLSA and directly close the current source if the corresponding CAM cell is \verb|MISMATCH|. We can also just extract the last several sense nodes close to the MLSA to do speculative sense if the CAM entry has hundreds of bits, which can reduce the wire routing effort in layout. Assuming the input data is random and every CAM entry has N bits, the probability of the \verb|MISMATCH| bits occurring in last n bits is $\frac{2^N-2^{N-n}+1}{2^N}$. As the example presented in Fig.~\ref{fig8}, extracting the last 3 bits from 10 bits CAM entry has 87.6\% probability closing the current source in advance when the CAM entry is \verb|MISMATCH| and test vector is evenly random.

To make sure the CSCD block can detect any current flow change under different matching cases, it's necessary to find the worst case for CSCD block, current flow change of which is smallest. Four different matching cases are analyzed here: a) All of the CAM entries are \verb|MATCH|, which means all of the MLSA can adaptively close current source by feedback control mechanism. b) All of the CAM entries have \verb|MISMATCH| bits in the last three CAM cells, then all of the MLSA can close current source by speculative sense mechanism before the request signal arrives and the current flow change in this case is only because of the charging current in dummy CAM entry. c) The \verb|MISMATCH| bits in all of the CAM entries only occur in the first eight CAM cells. The direct current in the pull-down path from VDD to GND exits in every CAM entry and can't be closed by speculative sense mechanism. d) Random matching cases. The current flow change is smallest when all of the CAM entries have \verb|MISMATCH| bits in the last three CAM cells based on the simulation results. Although it barely occurs in real application, it's important to make sure CSCD block has enough margin to generate the acknowledge signal in this case.

\subsection{Timing Analysis}
The CSCD block eliminates the second timing constraint introduced in section II and acknowledges the HS block after the CAM array finishes searching operation. In addition to the first timing constraint, which is that the request signal should be later than the valid data, there are two timing constraints related to HS circuits in CSCD block. One is minimum pulse width of clock signal, which corresponds to the duration of voltage \verb|Vs| change and the time interval of two search operations.  As introduced above, the minimum duration of voltage \verb|Vs| change occur in the case that all of the CAM entries have \verb|MISMATCH| bits in the last three CAM cells. The pulse width of current sensing circuits output is much longer than the minimum pulse width of clock signal even if in this case. The time interval of two search operations is also longer than the minimum pulse width of clock signal in practice. The other timing constraint is the minimum pulse width of reset signal, which is also simple to satisfy. The reason is same as above.

\subsection{Experimental results and discussion}
\begin{figure}
\centerline{\includegraphics[width=1\linewidth]{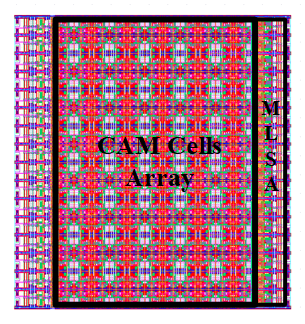}}
\caption{Layout of a CAM array with 16 CAM entries.}
\label{fig9}
\end{figure}

Evaluations are now presented for the new asynchronous CAM architecture. Results are obtained for two different sizes of full customized asynchronous CAM arrays with 16 CAM entries and 512 CAM entries respectively which are mapped to a 22FDX process FDSOI library. Each of proposed CAM architecture is compared to the conventional asynchronous CAM architecture in~\cite{Moradi_etal18} as the baseline architecture without CSCD, feedback control and speculative sense in terms of performance, power and area. Each CAM entry has 11 bits. The CAM array with 16 CAM entries is shown in Fig.~\ref{fig9}. Both CAM architectures use the same four-phase HS block. We carefully add dummy cells on the request signal path to make that the request signal has slightly higher capacitance load than SL and SLB, which is to satisfy the first timing constraint introduced in section II. Based on multiple Monte Carlo simulation results, the size of \verb|PD| transistors in dummy CAM entry are chosen to be 20\% larger than the \verb|PD| transistors in other CAM entries, which makes the charging speed of the dummy CAM entry slower than other \verb|MATCH| CAM entries. The 8bits configurable delay line is used in conventional asynchronous CAM architecture to fine tune the configurable delay. We start from 0 delay and increase it incrementally until there is no error signal, which is usually 30\% higher than the delay from request signal generation to dummy CAM entry output. The sense nodes in the last three CAM cells are extracted for speculative sense mechanism since they are closer to MLSA.

\subsubsection*{Cycle time}
\begin{figure}
\centerline{\includegraphics[width=1\linewidth]{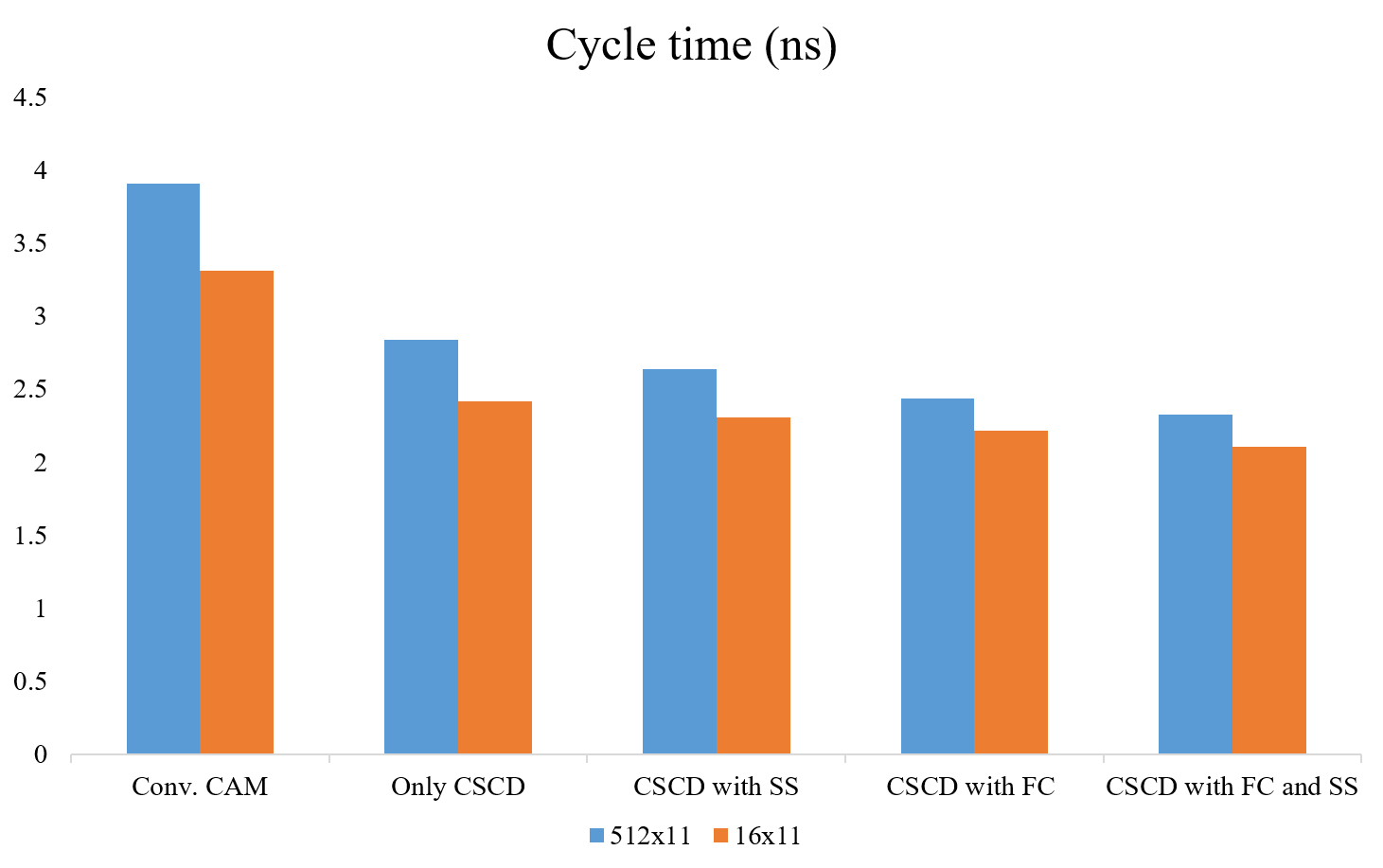}}
\caption{Average search cycle time.}
\label{fig11}
\end{figure}

Different from the evaluation for arbiters introduced in section III, we care more about the throughput of CAM architecture since the configurable delay line results large latency of asserting and deasserting stage in four-phase handshake. The average cycle time comparison between the conventional CAM architecture, proposed CAM architecture only with CSCD, proposed CAM architecture with feedback control or speculative sense and the complete proposed CAM architecture is shown in Fig.~\ref{fig11}, which can be directly translated to throughput performance. The average cycle time is calculated by multiple times random searching operation in the typical operating condition, which means the data input and the initial data content stored in the CAM array are random but the same for different CAM architectures. The complete proposed CAM architecture shows improvement for both 16x11 and 512x11 design points: 35.5\% and 40.4\%, respectively, over the conventional design. The higher performance improvement can be got in larger CAM array since the configurable delay line has to have higher delay as the CAM array size increases. In contrast, CSCD is not affected a lot by this, which provides a high acknowledge signal immediately when it detects that the CAM array finishes the searching operation. The result validates the benefits of CSCD. An interesting property of CSCD is that it also benefits from the feedback control and speculative sense as presented in Fig.~\ref{fig11}, since terminating charging earlier makes the current go back to zero in advance, which reduces the delay of providing acknowledge signal from CSCD block. More importantly, assuming that a configurable delay line always has higher delay than the searching operation is not robust because of device mismatch. Asynchronous CAM architecture with CSCD can eliminate the trade-off between performance and robustness introduced in Section II.

\subsubsection*{Area}
Post-layout areas are compared for the baseline vs. new   CAM architecture, at both design points. The final layout area is estimated by summing up the all of cells areas, including the CSCD and HS block. For 16x11 CAM architecture, the baseline design has an area of 225.3\,${\mu m}^2$, while the new CAM architecture occupies 245.5\,${\mu m}^2$. A 8.9\% increase of area is observed for 16x10 CAM architecture. The area increase is because of the CSCD block and the OR gate in MLSA , but there is no area increase in CAM cell even if we add extra one pin, which is important for scaling up the CAM array size. Such as for 512$\times$11 CAM architecture, the baseline and new designs have an area of 7242.1\,${\mu m}^2$ and 7620.6\,${\mu m}^2$, respectively. The area overhead increase of the new approach becomes less: only 5.2\%. 

\subsubsection*{Energy consumption}
This section reports the average energy consumption of both CAM architectures at 512x11 design point when all of CAM entries are \verb|MATCH|, all of CAM entries are \verb|MISMATCH| and random data searching. Although the first two extreme cases barely occur in neuromorphic processors, they are considered here since we want to specify the different power saving by different mechanisms. The \verb|MISMATCH| bits are distributed randomly in 11 bits CAM entry.

\begin{figure}
\centerline{\includegraphics[width=1\linewidth]{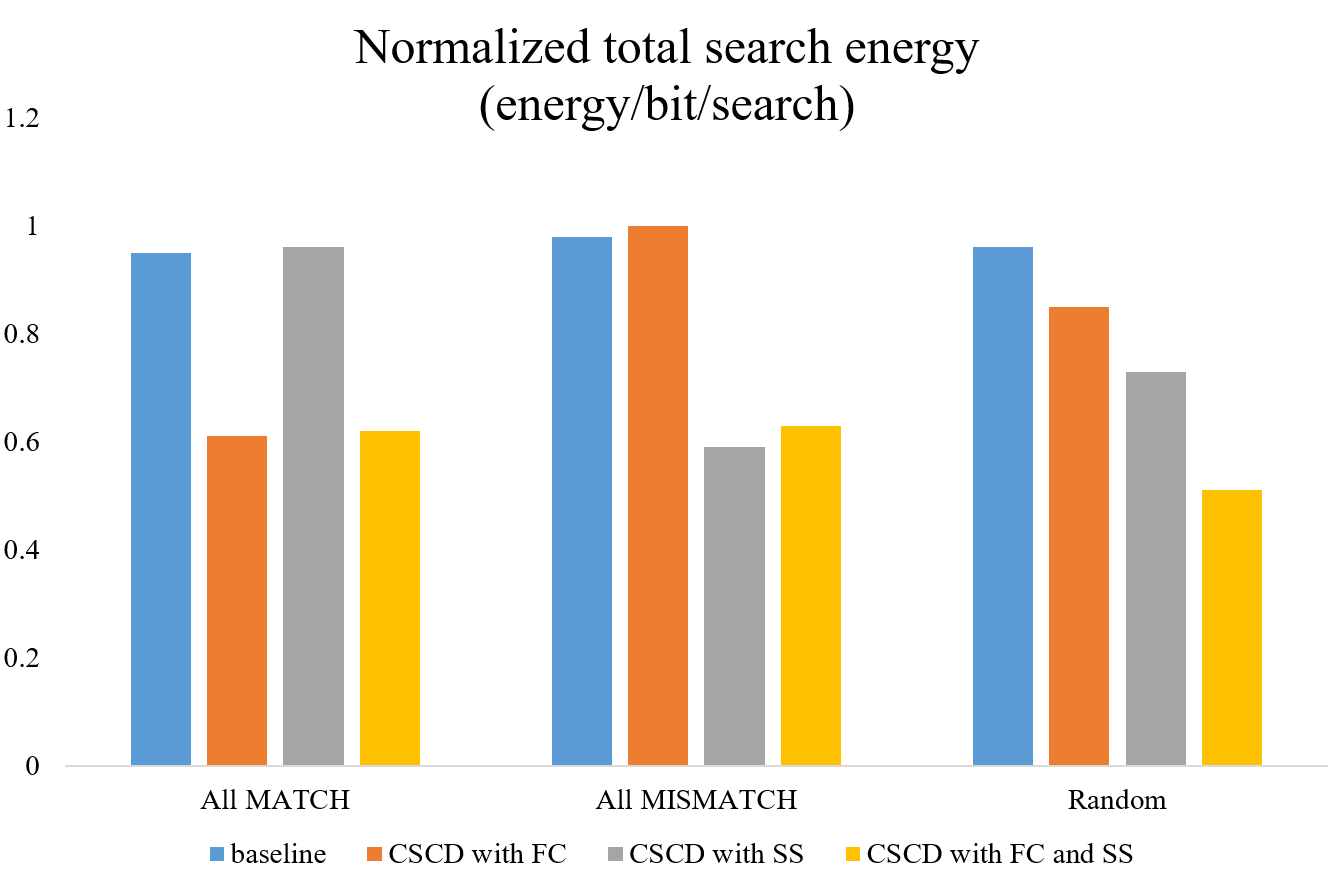}}
\caption{Normalized average search energy.}
\label{fig12}
\end{figure}

As shown in Fig.~\ref{fig12}, only feedback control and CSCD contribute to energy saving when all of CAM entries are \verb|MATCH|, which turns out to be 35.8\% lower than the baseline architecture. When considering all \verb|MISMATCH| case, the new CAM architecture taking advantage of speculative sense shows 40.2\% energy reduction. The newly designed CAM architecture based on CSCD, combined with the feedback control and speculative sense, results in 46.7\% energy saving when the CAM Array is provided random data input, which is the most energy efficient design choice. 

\section{Conclusion}
\label{sec:conclusion}

We proposed a HAT arbitration architecture to meet the demands of low-cost core interfaces for multi-core neuromorphic processors. In particular, we presented the encoding pipeline circuits in the core output interface and novel asynchronous CAM circuits based on a CSCD block, used in the core input interface.
We showed how the latency of new arbitration architecture is reduced by a factor up to 78.3\%, for sparse event operations, with a lower area cost compared to alternative state-of-the-art arbitration architectures.
The proposed asynchronous CAM architecture achieves a 40.4\% increase in throughput by CSCD, and a 46.7\% energy saving, due to feedback control and speculative sense mechanisms.
The current sensing circuits with lower sensing latency and power consumption is the direction to explore in the future, such as using current-mirror amplifier with local positive feedback to reduce the latency. Another challenge is to design suitable power rails for digital CAM array and analog CSCD circuits, which should minimize cross-talk and area cost. 

\section*{Acknowledgment}
The authors would like to thank Davide Bertozzi and Steven M. Nowick for the training of asynchronous circuits design flow and the insight of robust asynchronous memory and thank Tugba Demirci for the guidance on CAM circuit design.

\printbibliography

\end{document}